# High Temperature Ferromagnetism in Epitaxial Monolayers of Co-doped $Fe_5GeTe_2$

Jules Courtin,[1] Fatima Ibrahim,[1] Davide Benettin,[1] Djordje Dosenovic,[3] Roberto Sant,[2] Pâmella Vasconcelos Borges Pinho,[2] Vincent Polewczyk,[1] Alain Marty,[1] Isabelle Gomes de Moraes,[1] Matthieu Jamet,[1] Denis Jalabert,[3] Fadi Choueikani,[4] Philippe Ohresser,[4] Nicholas B. Brookes,[2] Hanako Okuno,[3] Mairbek Chshiev,[1,5] and Frédéric Bonell[1]*

*[1]University Grenoble Alpes, CNRS, CEA, IRIG-Spintec, F-38000 Grenoble, France*
*[2]European Synchrotron Radiation Facility, B.P. 220, 38043 Grenoble Cedex, France*
*[3]University Grenoble Alpes, CEA, IRIG-MEM, F-38000 Grenoble, France*
*[4]Synchrotron SOLEIL, L'Orme des Merisiers, 91192 Saint-Aubin, Gif-sur-Yvette, France*
*[5]Institut Universitaire de France (IUF), 75231 Paris, France*

Magnetic van der Waals materials have mainly been investigated in their bulk form or as few-layers flakes. Due to the challenges in producing atomically thin films, only a few have been isolated as monolayers, which typically exhibit long-range magnetic order below 150 K. In this work, we use molecular beam epitaxy to synthesize Co-doped $Fe_5GeTe_2$, achieving precise control over both thickness and composition. We demonstrate ferromagnetism well above room temperature in multilayer samples and present clear evidence of ferromagnetic ordering in monolayers up to ~200 K. The changes in Curie temperature and magnetic anisotropy with composition exhibit similar trends in both monolayers and thicker films, indicating that the magnetic properties are primarily governed by intralayer magnetic interactions. Through element-specific X-ray magnetic circular dichroism and density functional theory, we identify the substitution site of Co dopants and reveal the mechanism behind the Curie temperature enhancement induced by Co doping. Our findings suggest that, despite their weak magnetic moment, Co dopants strengthen the magnetic moments on neighboring Fe atoms and enhance the intralayer ferromagnetic exchange interactions.

Van der Waals (vdW) magnets have recently emerged as excellent systems for exploring low-dimensional magnetism and for the development of spin devices in which interfacial effects are significantly enhanced and controlled through proximity or external stimuli [1-3]. Notably, their vdW surface termination enables the stabilization atomically thin layers in which long-range magnetic order may persist, provided the system possesses a finite magnetic anisotropy. However, only a few vdW magnets have been isolated as truly two-dimensional layers. Those monolayers (MLs) generally exhibit rather low magnetic transition temperatures, 15 K in $MnBi_2Te_4$ [4], below 50 K in Cr trihalides [5-8], 118 K in $FePS_3$ [9], between 20 K and 130 K in $Fe_3GeTe_2$ [10-12], 146 K in CrSBr [13].

$Fe_NGeTe_2$ (N = 3-5) and $Fe_3GaTe_2$ compounds stand out as rare examples that exhibit itinerant magnetism, which is of great interest for spintronics [14,15]. This leads to strong, direct exchange interactions and relatively high Curie temperatures ($T_C$) that can be tuned using external factors such as electric fields [11], pressure [16], light [17], intercalation [18,19], or chemical substitution [20-23]. In $Fe_NGeTe_2$, increasing the Fe concentration from N = 3 to N = 5 has been shown to notably boost the magnetization and raise $T_C$ in bulk materials from approximately 230 K to 310 K [24-29]. However, the magnetic ground state of $Fe_5GeTe_2$ (F5GT) is particularly complex, due to the high sensitivity of the magnetic order to crystal characteristics – such as intralayer order [30-32], vdW stacking arrangements [25,33,34], self-intercalation [35,36] – and to chemical composition, including Fe vacancies [24,27,28,31].

Efforts to tune the magnetism of F5GT have included partial substitution of Fe with Co [20-22] or Ni [23,33]. These substitutions have led to increased $T_C$ values well above room temperature in bulk compounds (~360 K with 25% Co and ~400 K with 30% Ni). Ni doping induces significant structural distortions that correlate with changes in the magnetic behavior [23]. In contrast, Co doping has a subtler impact on lattice parameters but markedly alters magnetic ordering. Low Co concentrations (< 20%) enhance the ferromagnetic coupling, whereas at higher Co content (40–50%), both antiferromagnetic and ferromagnetic phases have been observed, linked to different vdW stacking sequences (AA vs AA′) [20,22]. Additionally, uncertainty remains regarding the precise location of Co dopants within the layers [20,22].

These complexities underscore the need for detailed microscopic understanding of both the structure and magnetic interactions in these materials. To date, most studies of vdW magnets have relied on bulk crystals or few-layer nanoflakes. $(Fe_{1-x}Co_x)_5GeTe_2$ (FCGT) has not been studied in the strictly two-dimensional regime, primarily due to the challenge of synthesizing samples with precise thickness control. In this Letter, we demonstrate large-area synthesis of FCGT films via molecular beam epitaxy (MBE), achieving precise control over both composition and thickness. We report robust ferromagnetism above room temperature in multilayer samples, and provide clear evidence of ferromagnetic order in MLs at temperatures up to ~200 K. Through a combination of element- and surface-sensitive X-ray magnetic spectroscopy and first-principles calculations, we determine the atomic positions of Co dopants and reveal their influence on intralayer magnetic interactions.

*Contact author: frederic.bonell@cea.fr

Our FCGT films were grown using MBE by coevaporation of Fe, Co, Ge, and Te. The elemental flux ratios were set to Co:(Fe+Co) = $x$, (Fe+Co):Ge = 5, and Te:Ge ≈ 3. Films with a thickness of 12 MLs were grown on $Al_2O_3$(0001) (sapphire) substrates under conditions similar to those described in [28,32] (see also Supplemental Material [37]), and subsequently capped with a 3-nm-thick Al layer that naturally oxidized upon air exposure. Due to a significant lattice mismatch with sapphire (-15%), FCGT growth typically begins with a disordered first ML that crystallizes during the deposition of the second layer, leaving a thin (6 Å) amorphous interfacial layer [28]. To achieve well-ordered ML films, 1-ML-thick FCGT was instead grown on lattice-matched Ge(111) substrates, with a misfit lower than 1.8% (hexagonal parameters $a_{FGT}$ = 4.07 Å, see Supplemental Material [37], and $a_{Ge}$ = 4.00 Å). These MLs were capped with a 3-nm-thick $CaF_2$ layer.

Structural and compositional characterizations were carried out using X-ray diffraction (XRD), scanning transmission electron microscopy (STEM), electron energy loss spectroscopy (EELS) and Rutherford backscattering spectrometry (RBS). Magnetic properties were investigated using the magneto-optical Kerr effect (MOKE) in longitudinal geometry and X-ray magnetic circular dichroism (XMCD) in total electron yield mode. The XMCD measurements were performed at the ID32 beamline of ESRF [42] and the DEIMOS beamline of SOLEIL synchrotron [43].

We first examine the 12-ML-thick FCGT films grown on sapphire. The θ-2θ XRD scans shown in Fig. 1b reveal a single set of Bragg reflections, confirming that the films are single-phase with uniform crystalline quality. Prominent Laue oscillations around the (003), (006), and (009) reflections confirm the film thickness and further indicate smooth interfaces and coherent crystal ordering throughout the film thickness. In F5GT ($x$ = 0), the measured interlayer spacing is 0.977 nm, corresponding to a $c$ lattice parameter of 2.7 nm, assuming a rhombohedral unit cell (with $c$ = 3 ML, see Fig. 1a). Upon Co substitution, a slight increase in interlayer distance is observed (+0.3% for $x$ = 0.25), along with a modest decrease in the in-plane lattice parameter (-0.3% for $x$ = 0.25) (see Supplemental Material [37]), a trend consistent with prior studies of bulk crystal [21]. EELS (Fig. 1c) shows spatially overlapping Fe and Co signals, confirming that Co replaces Fe within the layers without significant intercalation into the vdW gaps. RBS performed on a nominal $x$ = 0.5 FCGT sample (Fig. 1d) further confirms this substitution, yielding elemental ratios of Co:(Fe+Co) = 0.47±0.04, (Fe+Co):Ge = 4.86±0.39, and Te:Ge = 2.00±0.16.

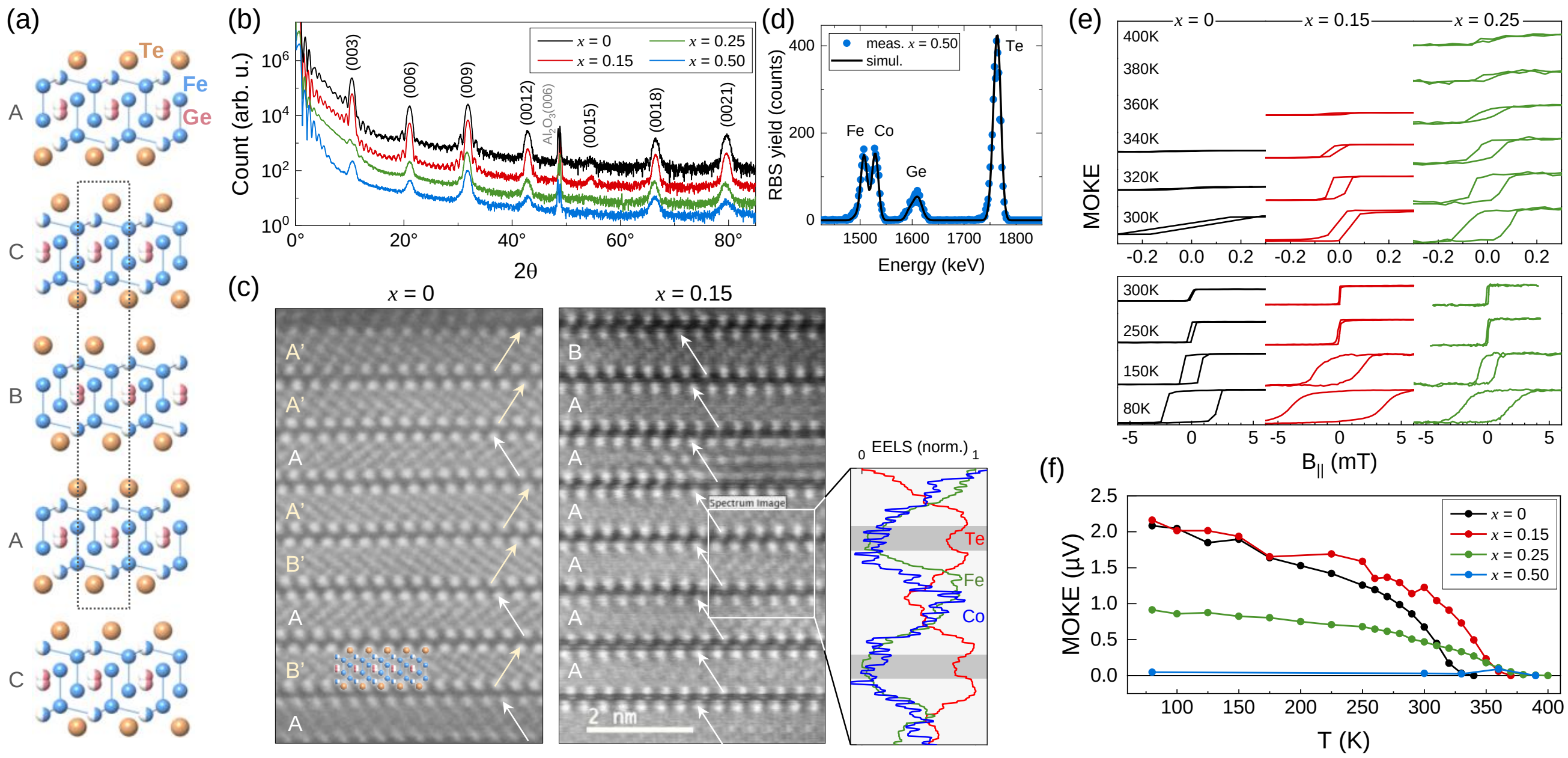

FIG 1. Properties of epitaxial 12-ML-thick FCGT layers as a function of the Co content $x$. (a) Illustration of the rhombohedral crystalline structure of F5GT. Half-filled atoms are symmetry-equivalent Fe1-Ge split sites. The dashed box highlights the unit cell with ABC stacking. (b) Out-of-plane θ-2θ X-ray diffraction scans. The peaks are labeled assuming a rhombohedral lattice. (c) STEM images for $x$ = 0 and $x$ = 0.15, and EELS spectra for $x$ = 0.15. Letters indicate the stacking order whereas arrows highlight the two twins (e.g. A or A'). (d) Experimental and fitted RBS spectra for the nominal composition $x$ = 0.50. (e) Longitudinal MOKE magnetic loops showing above-room-temperature ferromagnetism induced by Co doping. (f) Temperature dependence of the MOKE photodiode signal. The curves were shifted to zero at the highest measured temperature to account for small background offsets.

First-principles calculations predict that FCGT adopts a rhombohedral unit cell with ABC vdW stacking ($R\overline{3}m$ space group) for $x \leq 0.20$, transitioning to a primitive AA stacking sequence ($P\overline{3}m1$ space group) at $x = 0.60$ [20]. Experimentally, ABC stacking is typically observed in bulk F5GT, though often interrupted by AA stacking faults [24,25,34]. The AA stacking was shown to reinforce the $T_C$ and to favor in-plane alignment of the spin moments [34]. For $x \approx 0.5$, both AA and AA' stackings have been reported, where A′ corresponds to an in-plane 60° rotation of the A layer [22]. Fig. 1c displays representative STEM cross-sections for our films with $x = 0$ and $x = 0.15$. The F5GT film ($x = 0$) exhibits significant stacking disorder, with a high occurrence of AA stacking faults and rotational twins (A' or B'). Interestingly, moderate Co substitution ($x = 0.15$) clearly improves the stacking order, with the AA sequence becoming dominant and twinning largely reduced (see Supplemental Material for additional STEM and XRD characterization [37]).

Longitudinal MOKE measurements are presented in Figs. 1e,f. For $x \leq 0.25$, the films clearly display ferromagnetic behavior with easy-plane magnetization (i.e. the effective magnetic anisotropy is negative, as reported in [28]) and low coercive fields of a few mT at low temperatures (Fig. 1e). The coercivity peaks at $x = 0.15$, potentially indicating enhanced (negative) magnetic anisotropy at this composition. Co doping also increases $T_C$ by approximately 15%, from 320 K in FGT to ~360 K in FCGT (Fig. 1f), in line with earlier studies on bulk crystals [20,21] (see Supplemental Material for Arrott plot analysis [37]). The MOKE signal amplitude remains relatively stable up to $x = 0.15$ but drops sharply and disappears entirely for $x = 0.50$ (Fig. 1f). This reduction in signal suggests a loss of net magnetization at high Co content, possibly due to antiferromagnetic interlayer coupling or reduced magnetic moment within individual layers.

We now focus on FCGT MLs grown on Ge(111) substrates. STEM images (Fig. 2a) reveal continuous films with good thickness uniformity over tens of nanometers, as well as atomically sharp interfaces between FCGT and both the Ge substrate and the $CaF_2$ capping layer. The streaky and anisotropic reflection high-energy electron diffraction (RHEED) patterns (Fig. 2b) further confirm the epitaxy and high crystalline quality of the MLs across the full range of compositions studied. To assess the chemical composition, Fig. 2c presents the ratio of integrated X-ray absorption spectroscopy (XAS) white lines at the Co and Fe L edges, demonstrating proportionality to the nominal Co concentration $x/(1\text{-}x)$, which indicates efficient substitution of Co into F5GT at the ML scale.

Magnetic properties were investigated using XAS combined with XMCD measurements, performed with the experimental geometry illustrated in Fig. 3a, where a magnetic field $B$ is applied along the photon beam direction at an angle θ relative to the surface normal. Representative XAS/XMCD spectra measured at 20 K under a 5 T field in normal incidence (θ = 0°) are shown in Fig. 3c for the Fe and Co $L_{2,3}$ edges (see Supplemental Material for data analysis details [37]). The XAS spectra exhibit metallic-like line shapes regardless of Co content, accompanied by a strong XMCD signal, which reaches up to 30% of the XAS intensity at the Fe $L_3$ edge for $x = 0$. The lack of fine structure or multiplet splitting in the spectra supports the itinerant character of the magnetic interactions in these materials and confirms the absence of significant chemical interactions at the interfaces. As seen in Fig. 3b,c, the XMCD signals at Fe and Co edges share the same sign and field dependence, indicating that Co dopants are ferromagnetically coupled to Fe.

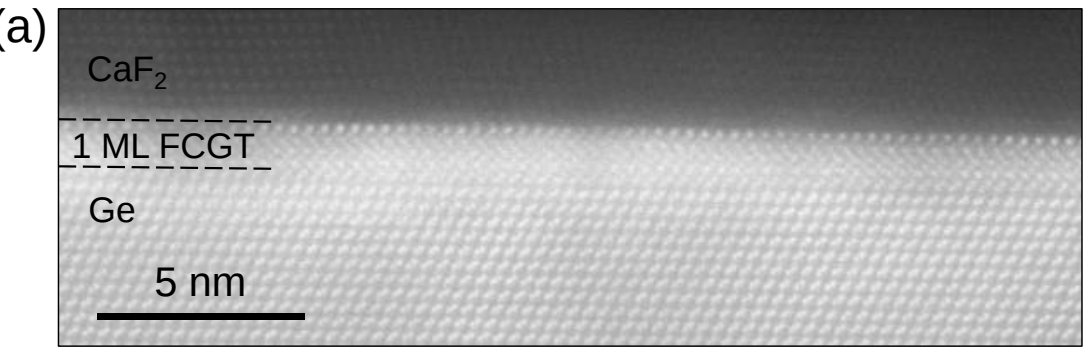

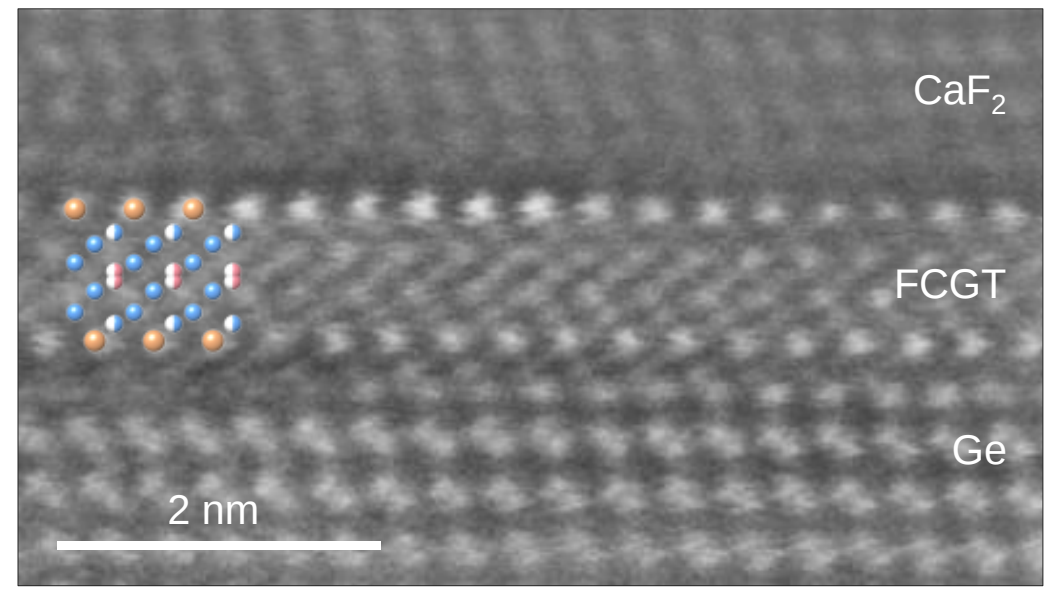

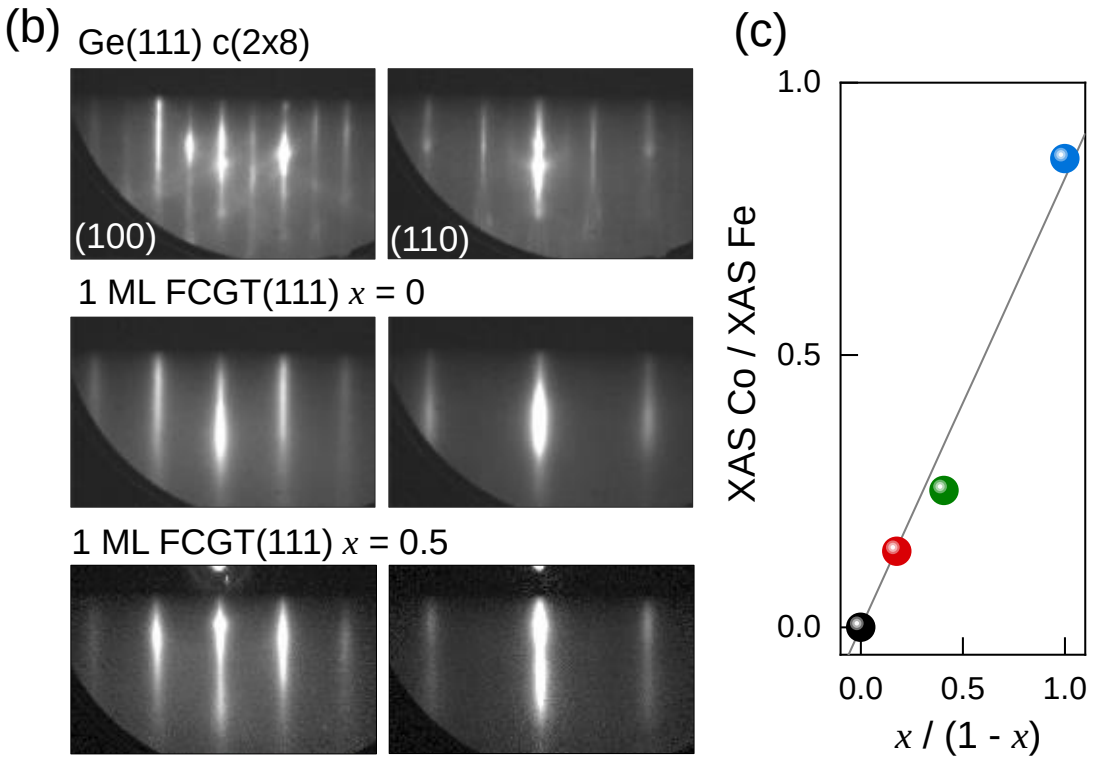

FIG 2. Epitaxial FCGT monolayers on Ge(111). (a) STEM images of a ML with $x = 0.15$. (b) RHEED patterns of the Ge(111) surface with its usual c(2×8) reconstruction and of MLs with different compositions. (c) Ratio of the integrated XAS white lines at Co and Fe L edges as a function of the nominal composition.

The strong XMCD signal at the Fe edge indicates the presence of long-range ferromagnetic order in FCGT MLs, further supported by the observation of hysteresis loops (Fig. 3d) and a non-monotonic temperature dependence of the XMCD intensity (Fig. 3e). At 20 K, clear magnetic hysteresis is observed for Co concentrations up to $x = 0.29$ (Fig. 3d). The effective magnetic anisotropy, close to zero at $x = 0$, becomes negative (easy-plane) with Co substitution and reaches a maximum at $x = 0.15$, in line with measurements on thicker films (see Supplemental Material [37] for a quantitative estimation of the magnetic anisotropy).

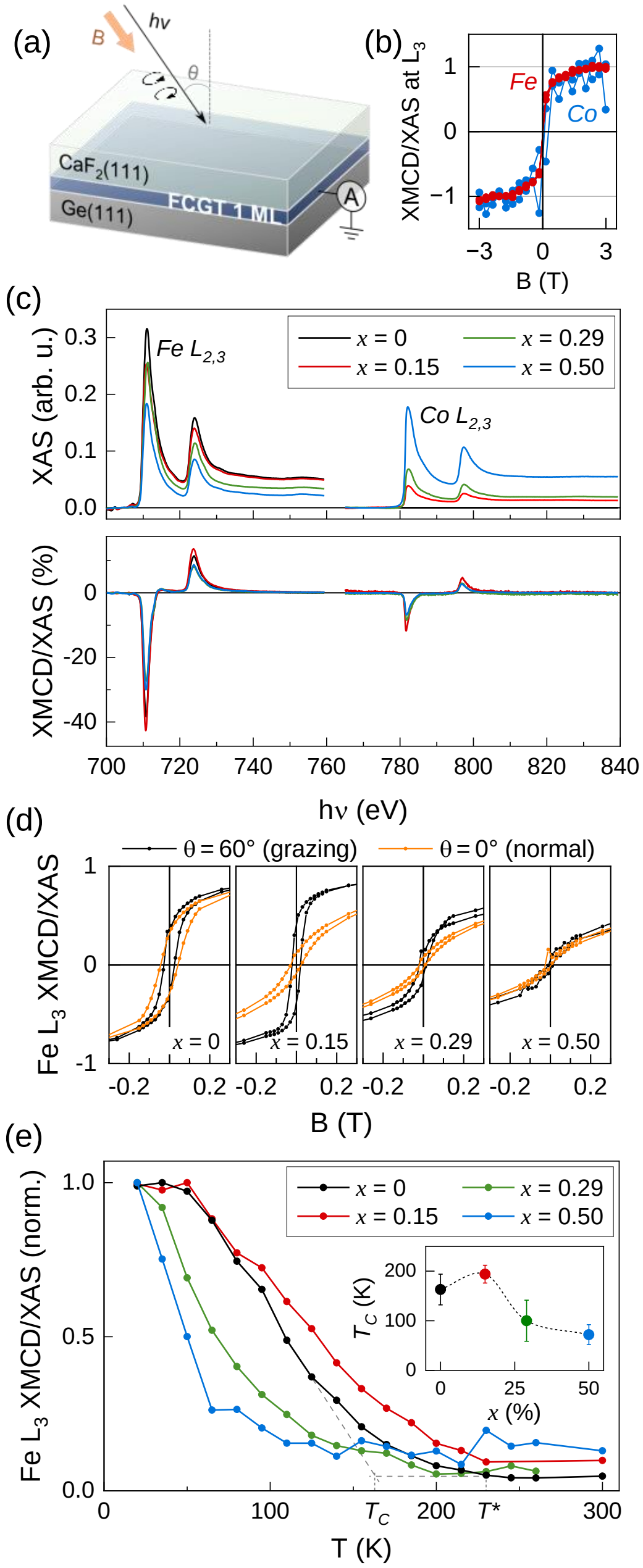

FIG 3. XAS/XMCD on FCGT monolayers. (a) Measurement geometry. (b) XMCD field loops at the Fe and Co $L_3$ edges, normalized to values measured at 3 T ($x$ = 0.29 / 20 K / θ = 60°). (c) XAS and XMCD spectra for various compositions (20 K / 5 T / θ = 0°). The XMCD is normalized to the $L_3$ XAS. (d) XMCD field loops at the Fe $L_3$ edge, normalized to values measured at 5 T (20 K). (e) Temperature dependence of the Fe $L_3$ XMCD near the magnetic remanence (θ = 60° / B = 50 mT after saturating under 5 T at 20 K). The signal is normalized to the XAS, then normalized to 1 at 20 K. Inset: estimated $T_C$ versus Co content. Error bars are $\pm(T^*-T_C)/2$ (see text).

In Fig. 3e, a ferromagnetic plateau is seen below 60 K, which shifts to lower temperatures as the Co content increases above $x = 0.15$ and disappears altogether at $x = 0.5$. This coincides with a reduction in coercivity and magnetic anisotropy that both vanish at this composition. Above the plateau, the XMCD signal gradually decreases with rising temperature, with the slowest decay observed at $x = 0.15$. A finite magnetic signal persists up to approximately $T^* \approx 230$ K for $x \leq 0.15$, and up to $T^* \approx 110$ K for $x = 0.5$ (the small ~1% offset above these temperatures is below our detection limit). Accurately determining $T_C$ is challenging because the temperature dependence of the XMCD does not follow a simple $(1-T/T_C)^\beta$ power law, as evidenced by the presence of an inflection point. We tentatively estimate $T_C$ as the turning point near the high temperature tail (Fig.3e), the latter being due to finite-size rounding of the transition [44,45]. The tail extension can be expressed as $(T^*/T_C - 1) \approx (a/\xi)^{1/\nu}$, where $\xi$ is the spin-spin correlation length, $a$ is of the order of the lattice parameter and $\nu$ (0.7-1.0) is a critical exponent [44,46,47]. From this expression, considering $\xi$ equals a few $a$, a broadening $T^*-T_C$ of a few tens of Kelvin is expected for a ML, which is generally consistent with our $T_C$ estimate. Other factors may contribute to the slow decay of the magnetization, namely, the small magnetic field $B = 50$ mT, applied to avoid ill-defined total electron yield signal in zero-field, and local thickness fluctuations.

Overall, these findings indicate that moderate Co doping strengthens the ferromagnetic order in FCGT MLs, while Co concentrations above 15% gradually reduce both $T_C$ and the magnetic anisotropy. Notably, this behavior closely mirrors that seen in thicker films. This strongly suggests that the magnetic properties of FCGT multilayers are primarily governed by intralayer magnetic interactions, while the modifications of the vdW stacking sequence induced by Co seem to play a secondary role. Note, however, that the finite-size broadening and lower $T_C$ relative to multilayers indicate that interlayer coupling also plays an important role in stabilizing the ferromagnetic order.

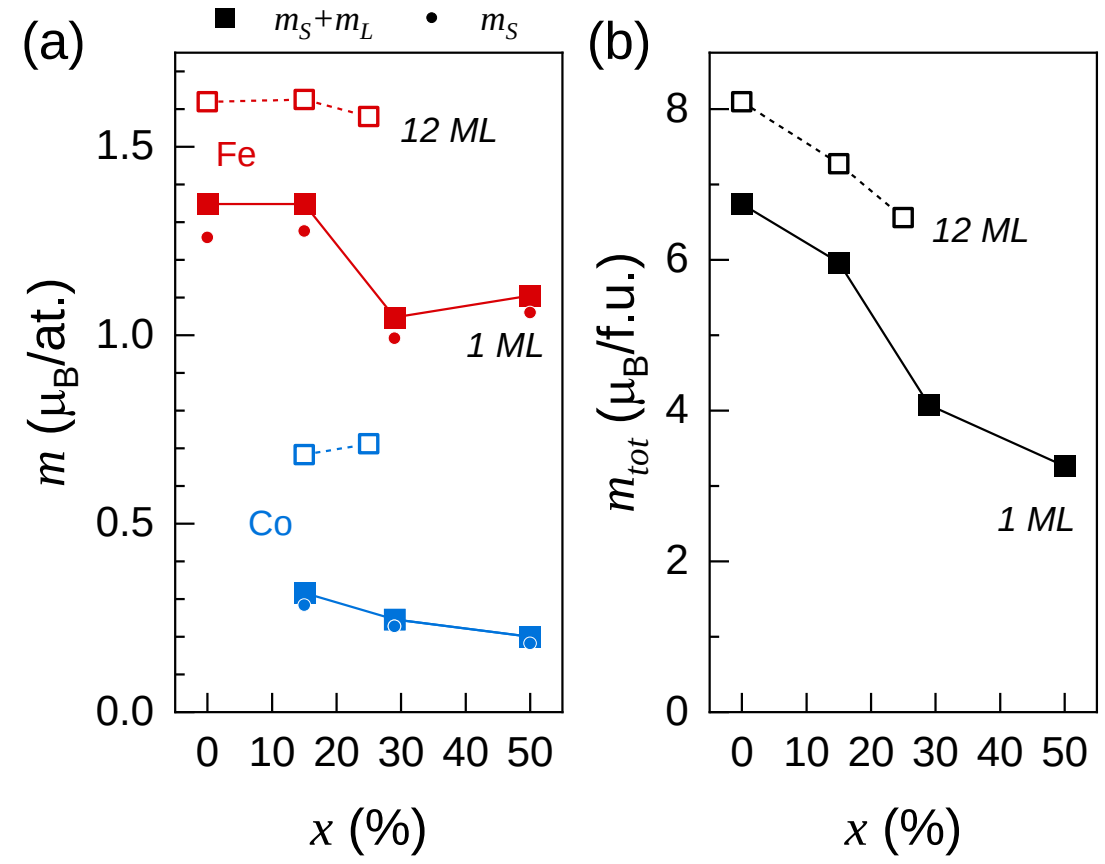

FIG 4. Measured magnetic moments in FCGT as a function of the composition. Empty and filled symbols are for 12 ML and 1 ML films, respectively. (a) Average magnetic moments on Fe and Co. (b) Total magnetic moment per formula unit (f.u.)

Fig. 4a presents the average magnetic moments of Fe and Co, as estimated using XMCD sum rules (see Supplemental Material [37]). The orbital-to-spin moment ratio remains below 0.1 across all samples, showing no significant dependence on composition. The key findings are as follows: (i) Increasing Co content results in a continuous decrease of the total magnetic moment per formula unit (Fig. 4b). (ii) Co atoms exhibit a very small magnetic moment, approximately 0.3 $\mu_B$/atom in MLs. (iii) The Fe magnetic moment remains large at $x = 0.15$, with indications that the spin moment may even increase slightly, but decreases with higher Co substitution. (iv) The overall trends are consistent between MLs and thicker films. These observations indicate that the variations in $T_C$ and magnetic anisotropy are more closely linked to the magnetic moment of Fe atoms than to that of Co, or to the overall average magnetic moment per unit cell.

First-principles calculations were carried out to better understand these experimental findings. Our analysis reveals that the ratio $m_{Co}/m_{Fe}$ strongly reflects the position of Co atoms within the unit cell (Fig. 5a, see also the Supplemental Material [37]). For a substitution level of one Co atom per formula unit (f.u.), i.e. $x = 0.20$, the energetically most favorable site for Co is Fe1 – nearest to the Te plane – followed by Fe4 (+0.18 eV/f.u.) and Fe3 (+0.30 eV/f.u.), consistent with earlier reports on bulk FCGT [20]. Experimentally, we measure a ratio $m_{Co}/m_{Fe} = 0.23$ at $x = 0.15$, which aligns closely with the theoretical prediction of 0.20. Substitution at other Fe sites would result in a ratio exceeding 0.5. This clearly indicates that in MLs, Co incorporates in the Fe1 sublattice with little site disorder. In contrast, a higher ratio of $m_{Co}/m_{Fe} = 0.42$ observed in multilayers suggests increased site disorder, with greater occupancy of Fe3 and Fe4 positions. An even larger ratio $m_{Co}/m_{Fe} = 0.56$ has been reported in a bilayer flake of FCGT with $x = 0.28$ [48], suggesting increased site disorder in bulk crystals. We also examined a scenario where Co alternates between the two symmetry-equivalent Fe1 sites near the top and bottom Te planes, yielding a theoretical ratio of 0.29 [37]. This specific kind of site disorder may contribute to the higher ratio in thicker films. Conversely, the lower ratio observed in MLs may reflect a preferential ordering of Co within one of the Fe1 sublattices – as reported for Co-rich FCGT [22] – thereby reducing the crystal symmetry to the polar *3m* point group. RHEED data supporting this scenario are provided in the Supplemental Material [37].

We now address the origin of the $T_C$ enhancement at moderate Co doping. Our calculations show that the total density of states at the Fermi level decreases with Co substitution, indicating that the $T_C$ increase cannot be attributed to the Stoner criterion alone. Therefore, site-specific magnetic interactions must be considered. Fig. 5b presents the spin-resolved charge redistribution upon substituting Fe1 with Co, where blue and red indicate electron accumulation and depletion, respectively. Co substitution introduces one extra electron per f.u., primarily filling the majority spin states of Co. This modifies the exchange interaction between site 1 and Fe2, flipping it from antiferromagnetic in the case of Fe1 to ferromagnetic in the case of Co1 (Fig. 5c, and Supplemental Material [37]). Despite this change, the local magnetic moment remains at 0.3 $\mu_B$, in excellent agreement with experimental data. Note that in the configuration where Co atoms occupy alternating Fe1 sites, the Fe1 site is ferromagnetically coupled to Fe2 and exhibits a higher moment of 1.9 $\mu_B$, which decreases to 0.6 $\mu_B$ upon Co doping. In actual films, some disorder likely exists in the Fe1 sublattice so an intermediate situation probably holds. Overall, Co incorporation primarily enhances the ferromagnetic coupling between Fe1 and Fe2 sublattices, which should naturally contribute to the observed $T_C$ increase. Additionally, we observe significant spin-dependent charge redistribution at the Fe2 site, raising its magnetic moment from 2.28 $\mu_B$ to 2.51 $\mu_B$. Notably, Fe2 hosts the largest magnetic moment and its bond lengths with Fe1 and Fe3, respectively 2.4 Å and 2.5 Å, are the shortest among all Fe pairs [23]. Therefore, the increased magnetic moment on Fe2 likely strengthens intralayer magnetic interactions and boosts $T_C$. We have checked that these findings remain valid when interfacing FCGT with Ge and $CaF_2$ [37]. Collectively, our measurements support the conclusion that Co substitution into the Fe1 sublattice enhances the magnetic properties of Fe, playing a central role in the observed $T_C$ enhancement.

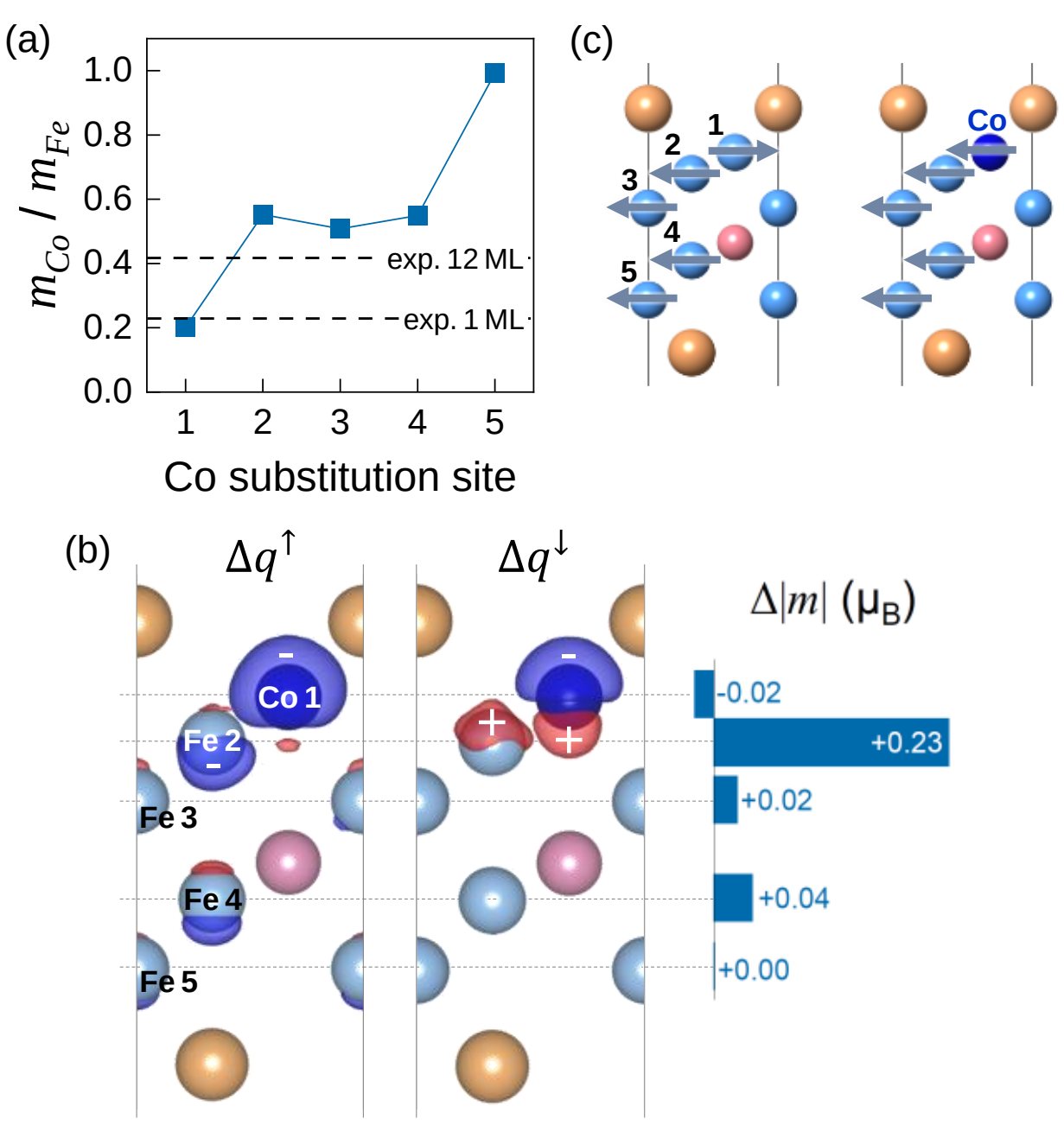

FIG 5. Ab initio calculations of FCGT monolayers with $x = 0.20$. (a) Ratio of the Co and Fe magnetic moments as a function of the Co substitution site for 1 ML on Ge(111). The horizontal dashed lines indicate measured values. (b) Charge variation in a freestanding ML upon substitution of Fe1 with Co for majority ($\Delta q^{\uparrow}$) and minority spins ($\Delta q^{\downarrow}$), taken at an isosurface value of $10^{-4}$ $e$/Å$^3$. The electron accumulation (depletion) is represented in blue (red). The bar diagram shows the variation of the magnetic moment on each site. (c) Ground state magnetic configurations of F5GT and FCGT.

In summary, we showed large-area epitaxy of FCGT with controlled composition and thickness down to the strictly two-dimensional regime. High temperature ferromagnetism (~200 K) was observed in MLs of FCGT, a rare example of two-dimensional vdW magnet with easy-plane magnetic anisotropy. The close agreement between the experimental and theoretical elemental magnetic moments allowed us to identify the Co substitution site and to clarify how Co may enhance the $T_C$ despite its weak magnetic moment. Remarkably, the impact of Co substitution on the $T_C$ and magnetic anisotropy shows a strong correspondence in multilayers and MLs, suggesting that the magnetic properties of FCGT are primarily determined by intralayer interactions. The large transition temperature of FCGT MLs, their sizable magnetic moment and the possibility to grow them by scalable epitaxy make them highly relevant to further investigations of low-dimensional magnetism and to their insertion in vdW heterostructures for spintronics.

## ACKNOWLEDGMENTS

This work was supported by the FLAG-ERA grant MNEMOSYN (ANR-21-GRF1-0005-01) and by the French ANR under projects ELMAX (Grant No. ANR-20-CE24-0015), SPINMAT (PEPR SPIN ANR-22-EXSP-0007), SPINTHEORY (PEPR SPIN ANR-22-EXSP-0009) and 2D-MAG (ESR/EQUIPEX+ ANR-21-ESRE-0025). The LANEF framework (No. ANR-10-LABX-0051) is acknowledged for its support with mutualized infrastructure. ESRF and SOLEIL are acknowledged for provision of synchrotron radiation facilities under proposals HC-5028 and 20220763, respectively. The RBS results were obtained at the AIFIRA facility, member of French accelerators network EMIR&A. AIFIRA is financially supported by the CNRS/IN2P3, l'Université de Bordeaux and the Région Nouvelle Aquitaine. We would like to thank the staff members of the AIFIRA facility, especially S. Vaubaillon and S. Sorieul.

## DATA AVAILABILITY

The data that support the findings of this article are not publicly available. The data are available from the authors upon reasonable request.